\RequirePackage{fix-cm}
\RequirePackage{rotating}

\documentclass[hidelinks,smallextended]{svjour3}

\smartqed

\usepackage{multirow}
\usepackage{balance}
\usepackage{hyperref}
\usepackage[authoryear]{natbib}

\hypersetup{
	colorlinks=true,
    citecolor=black,
    filecolor=black,
    linkcolor=black,
    urlcolor=black}

\begin{document}

\title{Do Citations and Readership Identify Seminal Publications?}

\author{Drahomira Herrmannova \and
		Robert M. Patton \and
        Petr Knoth \and
        Christopher G. Stahl
}

\institute{First author \at
           The Open University, Walton Hall, Milton Keynes, UK \\
           ORCID: 0000-0002-2730-1546 \\
           Phone: +1 (865) 253-9980 \\
           \email{drahomira.herrmannova@open.ac.uk} \\
           \emph{Current address:} Oak Ridge National Laboratory, Oak Ridge, TN, USA
           \and
           Second author \at
           Oak Ridge National Laboratory, Oak Ridge, TN, USA \\
           ORCID: 0000-0002-8101-0571
           \and
           Third author \at
           The Open University, Walton Hall, Milton Keynes, UK \\
           ORCID: 0000-0003-1161-7359
           \and 
           Fourth author \at
           Oak Ridge National Laboratory, Oak Ridge, TN, USA \\
           ORCID: 0000-0002-2070-1555
}

\date{Received: date / Accepted: date}

\maketitle

\begin{abstract}
In this paper, we show that citation counts work better than a random baseline (by a margin of 10\%) in distinguishing excellent research, while Mendeley reader counts don't work better than the baseline. Specifically, we study the potential of these metrics for distinguishing publications that caused a change in a research field from those that have not. The experiment has been conducted on a new dataset for bibliometric research called TrueImpactDataset. TrueImpactDataset is a collection of research publications of two types -- research papers which are considered seminal works in their area and papers which provide a literature review of a research area. We provide overview statistics of the dataset and propose to use it for validating research evaluation metrics. Using the dataset, we conduct a set of experiments to study how citation and reader counts perform in distinguishing these publication types, following the intuition that causing a change in a field signifies research contribution. We show that citation counts help in distinguishing research that strongly influenced later developments from works that predominantly discuss the current state of the art with a degree of accuracy (63\%, i.e. 10\% over the random baseline). In all setups, Mendeley reader counts perform worse than a random baseline.


\end{abstract}

\section{Introduction}

The question of how to accurately evaluate research outcomes is very difficult to answer, and despite decades of research, the problem is still largely unsolved. Under the current system published research should ideally receive a fair review by experts in the area and be given scores according to a generally accepted set of standards and rules. This process however often does not work \citep{smith2006peerreview,peplow2014peerreview}. This is largely due to the enormous and ever-growing number of papers being published every year, which was estimated to be over 1.5 million in 2008 \citep{jinha2010articles} and over 1.6 million in 2011 \citep{laakso2012anatomy}. As a consequence, researchers often resort to using metrics, such as journal and author impact factors and indexes, as a proxy to importance.

Recent years have seen the emergence of many new directions in this area, such as altmetrics \citep{galligan2013altmetrics}, webometrics \citep{almind1997informetric} and semantometrics \citep{knoth2014towards}. Despite the fact that research metrics attract much interest and new methods are constantly being developed, there exists no ground truth or reference dataset for assessing the usefulness of the existing and new methods. As a consequence, the authority of these methods is often established axiomatically.
For example, the two best-known metrics, the Journal Impact Factor (JIF) \citep{garfield1955citation} and the h-index \citep{hirsch2005index} were both proposed without empirical evidence demonstrating that they measure what they intend to measure. 

One possibility for validating these quantitative metrics is a comparison with peer review \citep{bornmann2013validation,waltman2014f1000}. Peer review, although not a perfect measure of quality in itself \citep{ioannidis_2014_make,begley_2015_reproducibility,teixeira_2015_problems}, is generally accepted as the most accurate research evaluation method \citep{bornmann2013validation}. However, there are some downsides to using peer review for evaluating research metrics, particularly the difficulty of obtaining data (unfortunately the only large dataset of peer review judgements known to us -- F1000Prime recommendations \citep{waltman2014f1000} -- isn't openly available). Furthermore, some researchers have called this method into question \citep{ricker2017letter}. The unavailability of a reference dataset complicates the development of new methods.

When talking about research evaluation and scientific impact and excellence, most people usually refer to the volume of change produced in a particular field (how much did a piece of work move the field forward), rather than referring to the educational (or other types of) impact generated. This is also the case for many national evaluation systems \citep{ref_2012_panel,tec_2013_quality,arc_2015_evaluation}. A characteristic example of the first type (publications which produced a high volume of change) are seminal publications, while literature reviews (surveys) are a typical example of the second type (publications generating different types of impact). Indeed, the definition of the word \textit{seminal} according to the Oxford Dictionary is ``strongly influencing later developments'' while the definition of the word \textit{review} is ``a report on or evaluation of a subject or past events'', which matches our understanding of the difference between these two types of papers. Hence, if one of the goals of research evaluation is recognising publications which contributed significantly to their field, seminal papers should perform better under such evaluation than literature reviews, which by definition do not generate a change in the field\footnote{With some exceptions, notably systematic reviews, which are a key practice in evidence-based medicine}. 

Therefore, we study how well the existing metrics discriminate between these two types of papers. Our results show that existing metrics help in distinguishing between seminal publications and literature reviews, albeit with room for improvement. We believe this is an important finding demonstrating more attention may need to be paid to publication type in research evaluation, especially as these two types of papers are weighted equally when used in research evaluation metrics such as in JIF \citep{mcveigh2009journal} and the h-index.

The work presented in this paper is conducted on a new dataset of seminal publications and literature reviews which we call TrueImpactDataset. This dataset was built from data collected in an online survey. We asked the respondents to provide two references from their research area -- a seminal publication and a literature review. We share this dataset with the research community\footnote{\url{http://trueimpactdataset.semantometrics.org/}} to help the development of new research evaluation metrics. The dataset consists of metadata (which include DOIs) of 314 research papers from different scientific disciplines -- 148 survey papers and 166 seminal papers. We use this dataset to study the predictive power of existing research evaluation metrics, namely citation counts and Mendeley reader counts, for identifying seminal publications. Furthermore, we discuss the parameters an ideal dataset for developing novel metrics should satisfy.

This paper is organized as follows. In Section \ref{sec:methodology} we describe our research question and how we aim to answer it. In Section \ref{sec:creation} we explain how the dataset was created. Section \ref{sec:analysis} presents some statistics of the dataset and Section \ref{sec:experiment} the results of an experiment in which we examine the value of citations and Mendeley reader counts in predicting the class of a paper. In Section \ref{sec:discussion} we discuss our findings and the properties an ideal evaluation set should have. Related work is presented in Section \ref{sec:relwork}.

\section{Methodology}
\label{sec:methodology}

This paper aims to answer the following research question: ``How well do citation and reader counts identify highly influential papers?'' Different approaches have been used in the past to try and answer this question. The approaches focused on studying the validity of citations for research evaluation can broadly be categorized into two groups. One group has focused on the unit of measurement -- citation itself, and has studied, for instance, the reasons for citing \citep{harwood2009interview} or not citing \citep{macroberts2010problems} specific papers, or the characteristics of citation, such as the placement \citep{bertin2016invariant}, and the context \citep{hu2015recurrence} of citations in text. The second group has concentrated on understanding what citations represent, for example by studying the characteristics of highly cited publications \citep{antonakis2014makes} and whether they correlate with external indicators \citep{bornmann2015does}. 

We adopt a slightly different method. A typical data analysis/statistics approach to answering the question above would be to test the metrics on a ranked set of papers and to express the success rate of these metrics using an evaluation measure such as precision and recall. However, to our knowledge, there exists no openly available ground truth or a reference dataset that could be used for establishing the validity of research evaluation metrics. While there was an attempt at creating such a dataset (Section \ref{sec:relwork}), this dataset wasn't openly shared and so cannot assist with this task. A similar dataset which has recently been used for this purpose also isn't openly available \citep{waltman2014f1000}. Because building such dataset would require significant time and resources (Section \ref{sec:discussion}) we were looking for an alternative approach for validating the metrics.

As mentioned in the introduction, when talking about evaluation of research outputs, an important dimension is the amount of change produced in a research area (how much was the area pushed forward thanks to a given piece of work) \citep{ref_2012_panel,tec_2013_quality,arc_2015_evaluation}. This amount of change has been discussed and studied from different perspectives \citep{Yan2012,knoth2014towards,Whalen2015,valenzuela2015identifying,Patton2016}. We were looking for a sample of research publications representing such work and we believe seminal research papers constitute such sample. To provide a clear comparison we were also interested in review publications (papers presenting a survey of a research area). While these papers are often highly cited \citep{Seglen1997,Aksnes2003} they usually don't present new original ideas. In this paper we study how well citation counts and Mendeley reader counts distinguish between these two types of papers.

To our knowledge, there currently isn't any dataset which would categorize papers into these two categories. We were therefore left with creating such dataset ourselves. We have employed an online survey for this task. The format of the survey, the number of collected responses and other details are presented in Section \ref{sec:collection}. In the following section (\ref{sec:analysis}) we analyze the dataset to understand whether it is suitable for our purposes.

In order to answer our research question, we have designed a simple experiment. We chose citation counts and Mendeley readership as representatives of bibliometrics and altmetrics, as these two measures are both well known and are being used as measures of impact of published research in many settings \citep{ref2014criteria,wilsdon2015metrictide}. We then classify the papers in the collected dataset into two classes (seminal, review) using two models, a model using the papers' citation counts and a model using their Mendeley readership (Section \ref{sec:experiment}). We show that the model using citation counts outperforms our baseline by a significant margin, while the model using readership doesn't perform better than the baseline.

\section{Dataset creation}
\label{sec:creation}

This section describes the dataset and the process used to create it. The dataset is publicly available for download\footnote{\label{note:truid}\url{http://trueimpactdataset.semantometrics.org}}.

\subsection{Initial data collection}
\label{sec:collection}

The goal was to create a collection of research publications consisting of two types of papers, seminal works, and literature reviews. We have used an online form to collect the references, which was composed of two sets of questions -- questions about the respondent's academic background (their discipline, seniority and publication record) and questions which asked for a reference to a seminal paper and to a literature review, both related to the respondent's discipline. We have used the latest Research Excellence Framework (REF) units of assessment \citep{refunits} as a list of disciplines when asking about the respondents' academic background because UK researchers are familiar with this classification.

The survey was sent to academic staff and research students from all faculties of the Open University (to 1,415 people in total). The reason why we contacted Open University researchers is because research at the Open University covers many disciplines, and because it is the largest university in the United Kingdom. We were therefore able to get a significant sample spanning multiple disciplines. Within three months we have received 184 responses (172 references to seminal papers and 157 to review papers), which represents a 13\% response rate. The survey questions and email invitation are available online together with the dataset\textsuperscript{\ref{note:truid}}.

To enable the respondents to send at least one reference, in case they weren't able to submit both, we made both answers optional. Ten respondents have only filled the questions related to their academic background but have not provided the references. We have removed these responses from the dataset which left us with 174 responses.

We didn't require the references to be in a specific format (e.g. a URL or DOI) to make it easier to complete the survey. The respondents were allowed to submit the references in any format they preferred (as a text, link, etc.). As a consequence, a few of the references were submitted in a format which made it impossible for us to identify the papers (e.g. ``Stockhammer (2004)''). We have removed these papers from the dataset. After removing empty and unidentifiable responses, we were left with 171 responses providing us with 166 seminal and 148 literature reviews.

\subsection{Additional metadata}
Once the survey was closed we have manually processed the data and collected the following information (by querying a search engine for the paper title and looking for a relevant page): a DOI, or a URL for papers for which we did not find a DOI, title, list of authors, year of publication, number of citations in Google Scholar and abstract. Where we had access to the full text, we have also downloaded the PDF. We were able to download 275 PDFs and 296 abstracts. Due to copyright restrictions, the PDFs are not part of the shared dataset\footnote{As there are Copyright Exceptions for text and data mining in some countries, such as in the UK, we are happy to provide the PDF documents for these purposes to researchers residing in these jurisdictions upon request.}. This collection process took a single person several hours a day for about a week.

To obtain readership data, we have used the DOIs, or title and year of publication for papers without a DOI, to query the Mendeley API\footnote{\url{http://dev.mendeley.com}}. We were mainly interested in the number of readers of each paper. The dataset contains a snapshot of the Mendeley metadata we were working with. We were able to find 141 out of the 166 seminal papers and 125 out of the 148 literature reviews in Mendeley.

Furthermore, using the Web of Science (WoS) API\footnote{\url{http://ipscience-help.thomsonreuters.com/wosWebServicesLite/WebServicesLiteOverviewGroup/Introduction.html}} we managed to retrieve additional information for the seminal and literature review papers indexed by WoS. We queried the WoS API using publication DOIs, if the document was in the system we obtained a full list of publications citing the paper in question and publications cited by the paper. This list included minimal metadata. In order to get full citation information, we queried the API for each individual (citing and cited) paper.

\section{Dataset analysis}
\label{sec:analysis}

To ensure the collected dataset is suitable for our task, we looked several statistics describing the dataset including statistics of publication age, distribution across disciplines and citation and readership statistics.

\subsection{Size}

The size of the dataset is presented in Table \ref{tab:size}. The row \textit{DOIs} shows the number of papers in the dataset for which we were able to find a DOI and the row \textit{DOIs in WoS} how many of these DOIs appear in the Web of Science database. The number of additional references which we collected using the WoS API is shown in the row \textit{Citing \& cited references}.

The rows \textit{Authors total} and \textit{Unique author names} show the total number of authors of all papers in the dataset and the number of unique author names. To count the unique names, we have compared the surname and all first name initials, in case of a match we consider the names to be the same (e.g. J. Adam Smith and John A. Smith will be counted as one unique name). The \textit{Unique author names} column doesn't show the number of disambiguated authors, but gives us an indication of how many of the author names repeat in the dataset.

\bgroup
\def\arraystretch{1.5}
\begin{table}[ht!]
    \centering
    \small
    \begin{tabular}{|l|r|}
        \hline
        Responses & 171 \\
        \hline
        Seminal papers & 166 \\
        \hline
        Review papers & 148 \\
        \hline
        Total papers & 314 \\
        \hline
        Seminal in Mendeley & 141 \\
        \hline
        Review in Mendeley & 125 \\
        \hline 
        Total in Mendeley & 266 \\
        \hline
        DOIs & 256 \\
        \hline
        DOIs in WoS & 110 \\
        \hline
        Authors total & 1334 \\
        \hline
        Unique author names & 1235 \\
        \hline
        Abstracts & 296 \\
        \hline
        Citing \& cited references & 19,401 \\
        \hline
    \end{tabular}
    \caption{Dataset size.}
    \label{tab:size}
\end{table}
\egroup

\subsection{Publication age}
\label{sec:age}

\begin{figure*}[ht!]
 	\centering
	\includegraphics[width=1.0\linewidth]{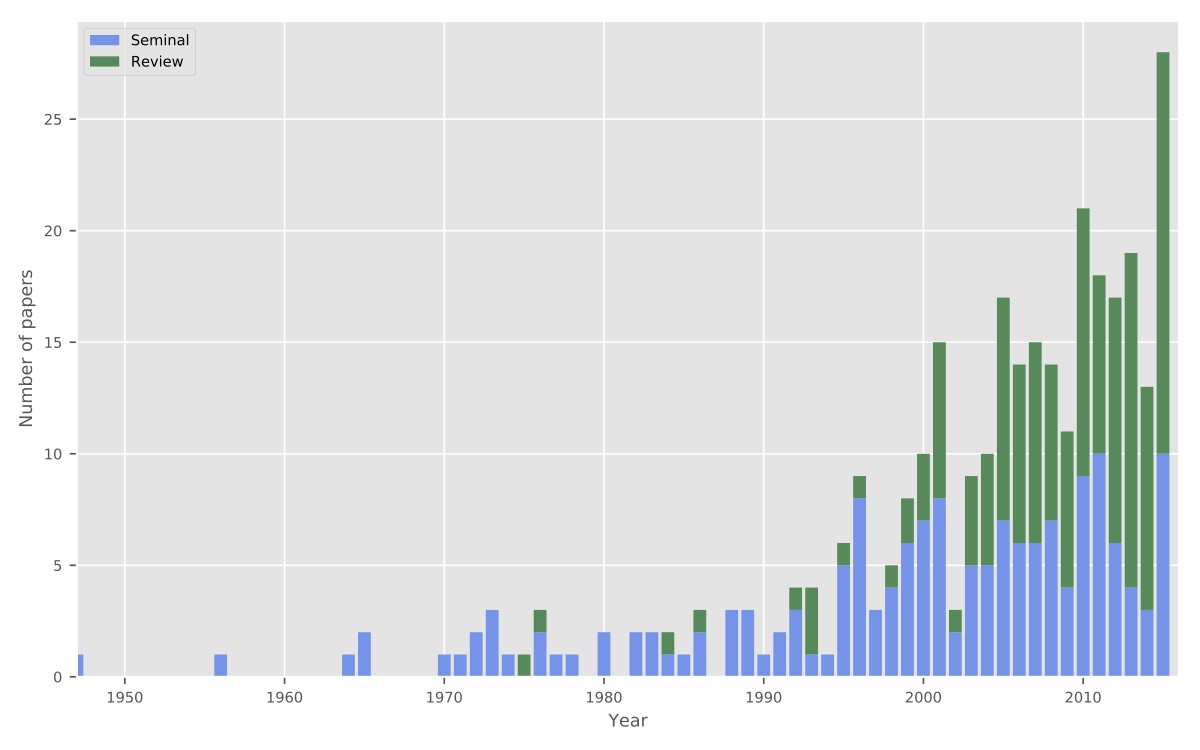}
    \caption{Histogram of publication years.}
    \label{fig:years_dist}
\end{figure*}

Figure \ref{fig:years_dist} shows a histogram of years of publication with literature reviews and seminal papers being distinguished by colour. Seminal papers in the dataset are on average about 9 years older than review papers. This shows literature reviews might age faster than seminal papers, which is consistent with our expectations. An explanation for this could be that literature reviews theoretically become outdated as soon as the first new piece of work is published after the publication of the review. Because the seminal papers are on average older this also means these papers had more time to attract citations. This is another reason to expect seminal papers to be distinguishable by citations and readership as features. Descriptive statistics of years of publication both sets are presented in Table \ref{tab:years}.

\bgroup
\def\arraystretch{1.5}
\begin{table}[ht!]
	\centering
    \small
    \begin{tabular}{|l||r|r|r|}
    	\hline
        & \textbf{Seminal} & \textbf{Review} & \textbf{Overall} \\
        \hline
        Mean & 1999 & 2008 & 2003 \\
        \hline
        Min & 1947 & 1975 & 1947 \\
        \hline
        Max & 2016 & 2016 & 2016 \\
        \hline
        25\% & 1995 & 2005 & 1999 \\
        \hline
        50\% (median) & 2002 & 2010 & 2006 \\
        \hline
        75\% & 2010 & 2013 & 2011 \\
        \hline
    \end{tabular}
    \caption{Descriptive statistics of publication age for both types of papers.}
	\label{tab:years}
\end{table}
\egroup

\subsection{Disciplines}

\begin{figure*}[ht!]
 	\centering
	\includegraphics[width=1.0\linewidth]{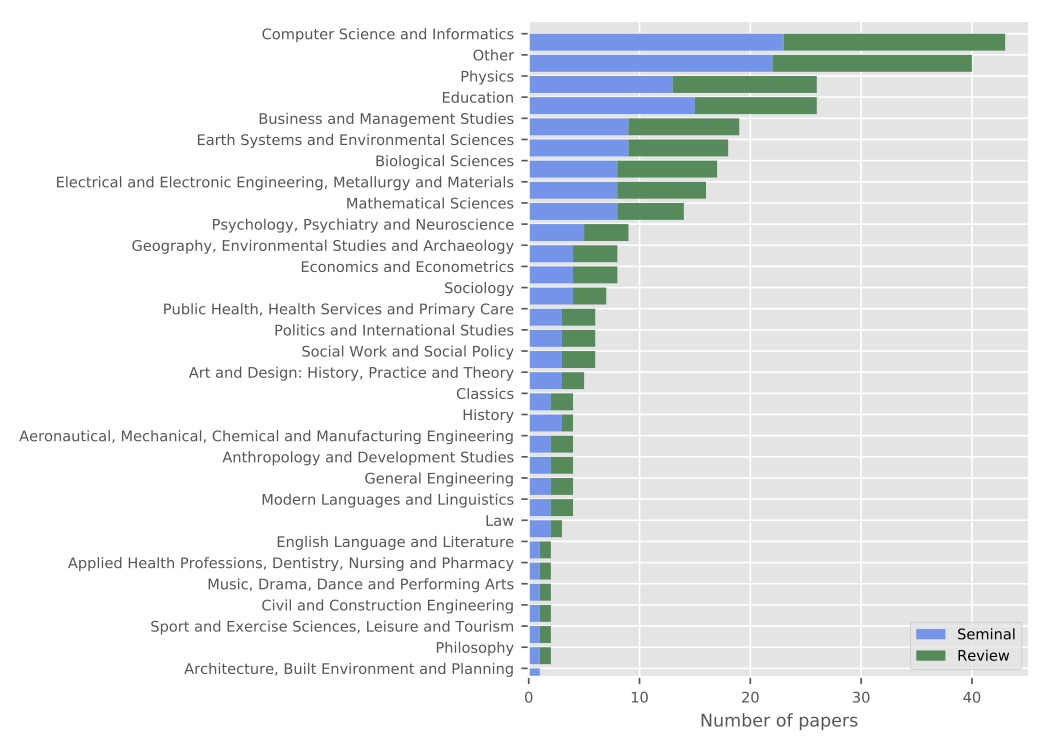}
    \caption{Histogram of publication disciplines.}
    \label{fig:disciplines}
\end{figure*}

Figure \ref{fig:disciplines} shows a histogram of papers per discipline. We have used the information we got about the respondents' academic background to assign papers to disciplines. The respondents have also provided a short description of the research area related to the two references (e.g. ``molecular neuroscience'', ``combinatorics'', etc.), however as these descriptions are more detailed and there is little overlap between them we haven't used these in our analysis. 

The distribution of papers per discipline is to a certain degree consistent with other studies, which have reported Computer Science and Physics to be among the larger disciplines in terms of number of publications, however, Medicine and Biology are typically reported to be the most productive \citep{Althouse2009,DAngelo2015}. The distribution is therefore probably more representative of size of faculties of the Open University than of productivity of scientific disciplines in general, however, we believe this does not influence our study.

When answering the questions about academic background, 22 respondents have selected ``Other'' instead of one of the listed disciplines, these 22 responses provided us with 40 papers in total. We looked at the detailed description of these 40 papers, 9 of them are related to astronomy (the descriptions provided were ``Binary stars'', ``Martian meteorites'', ``cosmochemistry'', ``Planetary sciences'' and ``planetology''), 4 could be classified as computer science (``virtual reality'' and ``Natural Language Understanding, Spoken Language Understanding''), the rest relate to different areas (e.g. ``Microbial degradation of plastic'' or ``MOOC'').

\subsection{Citations and readership}

The dataset contains two basic measures related to publication impact and utility -- citation counts, which we manually collected from Google Scholar, and the number of readers in Mendeley, which we gathered through the Mendeley API. We also had access to the number of citations in Web of Science and while we couldn't make these data available together with the dataset, we provide an analysis of the WoS citations and a comparison with the other two metrics.

Table \ref{tab:cit_reader_stats} shows basic statistics of Google Scholar citation counts and Mendeley readership of each paper in the dataset. We consider the readership of papers which we didn't find in Mendeley to be 0 (as papers are added to the Mendeley database by their readers). It is interesting to notice that while seminal papers are on average cited more than review papers, this is not the case for readership, in fact literature reviews attract more readers than seminal papers despite being on average younger (Section \ref{sec:age}). We believe this is an important finding as readership counts are being more and more frequently used as a measure of impact complementary to citations \citep{piwowar2013power,maflahi2016readership,priem2016altmetrics}. We believe the fact that literature reviews are more read than seminal papers, while being less cited, suggests that readership can be perceived more as a measure of popularity than importance.

\bgroup
\def\arraystretch{1.5}
\begin{table*}[ht!]
	\centering
    \small
    \begin{tabular}{|l||r|r|r||r|r|r|}
    	\hline
        & \multicolumn{3}{c||}{\textbf{Google Scholar citations}} & \multicolumn{3}{c|}{\textbf{Mendeley readership}} \\
        \hline
        & \textbf{Seminal} & \textbf{Review} & \textbf{Overall} & \textbf{Seminal} & \textbf{Review} & \textbf{Overall} \\
        \hline
        Mean & 2,458 & 519 & 1,544 & 240 & 368 & 306 \\
        \hline
        Std &  8,885 & 1,197 & 6,575 & 894 & 1,566 & 1,264 \\
        \hline
        Min & 0 & 0 & 0 & 0 & 0 & 0 \\
        \hline
        Max & 85,376 & 12,099 & 85,376 & 10,258 & 15,516 & 15,597 \\
        \hline
        25\% & 78 & 24 & 41 & 6 & 7 & 7 \\
        \hline
        50\% (median) & 249 & 109 & 194 & 45 & 42 & 46 \\
        \hline
        75\% & 1,302 & 596 & 845 & 166 & 145 & 165 \\
        \hline
    \end{tabular}
    \caption{Descriptive statistics of Google Scholar citation counts and of Mendeley readership.}
	\label{tab:cit_reader_stats}
\end{table*}
\egroup

Table \ref{tab:cit_gs_wos} shows a comparison of the number citations obtained from Google Scholar and from WoS. This table includes only those 110 papers (51 seminal and 59 survey papers) which appear in WoS. The higher citation numbers coming from Google Scholar are not surprising as Google Scholar's wider coverage of academic outputs is well known \citep{Harzing2016a,Harzing2016b}. This wider coverage is also demonstrated by the fact that we were able to find only 110 out of the 314 papers used in our study in WoS. 

\bgroup
\def\arraystretch{1.5}
\begin{table*}[ht!]
	\centering
    \small
    \begin{tabular}{|l||r|r|r||r|r|r|}
    	\hline
        & \multicolumn{3}{c||}{\textbf{Google Scholar}} & \multicolumn{3}{c|}{\textbf{Web of Science}} \\
        \hline
        & \textbf{Seminal} & \textbf{Review} & \textbf{Overall} & \textbf{Seminal} & \textbf{Review} & \textbf{Overall} \\
        \hline
        Mean & 814 & 429 & 607 & 523 & 255 & 379 \\
        \hline
        Std & 1,599 & 566 & 1,175 & 926 & 373 & 697 \\
        \hline 
        Min & 2 & 0 & 0 & 1 & 0 & 0 \\
        \hline
        Max & 8,246 & 2,446 & 8,246 & 4,753 & 1,709 & 4,753 \\
        \hline
        25\% & 102 & 43 & 59 & 46 & 25 & 33 \\
        \hline
        50\% (median) & 211 & 216 & 214 & 144 & 94 & 105 \\
        \hline
        75\% & 929 & 612 & 705 & 677 & 354 & 418 \\
        \hline
    \end{tabular}
    \caption{Descriptive statistics of citation counts acquired from Google Scholar and Web of Science. This table includes only those 110 papers, which we were able to find in WoS.}
	\label{tab:cit_gs_wos}
\end{table*}
\egroup

This low coverage provided by Web of Science can be seen as a problem, especially given the fact WoS misses some key seminal papers and overall misses more seminal papers that literature reviews. For example, a recent publication by Krizhevsky et. al. \citep{krizhevsky2012imagenet}, a seminal deep learning paper which has caused a shift in the area of artificial intelligence/computer vision, is missing in WoS, but has (at the time of writing this paper) attracted almost 8000 citations in GS since its publication in 2012. This problem isn't limited to WoS either, Scopus for example also does not index the publication, and while Mendeley does, most of the associated meta-data is inaccurate. The most probable reason for these exclusions is that the conference proceedings for this paper are not published through a major publisher but instead by the conference itself and self-hosted on their website. We believe this is an interesting point as it shows important seminal work isn't always published by the traditional routes of journals or known publishers. With the recent changes in scholarly communication towards Open Access, Open Science, Arxiv, self hosting, etc. the very definition of ``published'' no longer has a universal standard and we believe it is reasonable to expect that this will continue with higher frequency as the communities continue to change over time.

In order to compare whether the two databases rank papers similarly we have correlated the citation counts (see Table \ref{tab:cit_corr}). Both correlations are weaker for seminal papers, however this could be caused by the age difference between the two types of papers as the databases might have a lower coverage of older publications. Overall, both Pearson and Spearman correlations are otherwise strong. We believe this shows using citation data from these two databases will produce similar results.

\bgroup
\def\arraystretch{1.5}
\begin{table}[ht!]
	\centering
    \small
    \begin{tabular}{|l||r|r|}
    	\hline 
        & \textbf{Spearman} & \textbf{Pearson} \\
        \hline 
        Seminal & $0.8581, p \ll .001$ & $0.6775, p \ll .001$ \\
        \hline 
        Review & $0.9696, p \ll .001$ & $0.9588, p \ll .001$ \\
        \hline
        Overall & $0.9281, p \ll .001$ & $0.7254, p \ll .001$ \\
        \hline
    \end{tabular}
    \caption{Correlation between Google Scholar and Web of Science citation counts.}
	\label{tab:cit_corr}
\end{table}
\egroup

\section{Experiment \& Results}
\label{sec:experiment}


In this section, we present the results of the experiment the aim of which was to test whether citation or readership counts work as a discriminating factor for distinguishing seminal papers and literature reviews. These two measures, and especially citation counts, are frequently used  as proxies for scientific influence and quality. For example, citation counts are the basis for calculating JIF, where the calculation doesn't take into account the differences between types of research papers (pure research papers and literature reviews are both used as input with equal weight) \citep{jcr_source_data}. Amount of research contribution is often indicated as a dimension of research quality \citep{ref_2012_panel,tec_2013_quality,arc_2015_evaluation}. Thus, we study how well do these two types of papers distinguish between publications generating very different amounts of research contribution.

In order to test our hypothesis we use these two metrics to classify the papers into the two classes (seminal, review). As a baseline we use a model which classifies all papers as seminal, as that is the majority class. This baseline model achieves the accuracy of 52.87\%. We calculate accuracy as the proportion of correctly classified publications, or more formally:

\begin{equation}
	acc = \frac{TP + TN}{N}
\end{equation}

where the category \textit{seminal} is our positive class, $TP$ (true positives) is the number of items correctly labelled as belonging to the positive class, $TN$ (true negatives) is the number of items correctly labelled as not belonging to the positive class, and $N$ is the number of all items (publications).

Before running the experiments we first perform a statistical test to see whether the citation/readership distributions of seminal and review papers differ. We perform a one-tailed independent t-test with the null hypothesis stating that the means of the two groups are equal. The results we get are $p=0.0063$ for citations and $p=0.1666$ for reader counts. In case of citations, for a significance threshold of 1\% we reject the null hypothesis. Because we know the mean number of citations of the seminal papers is higher (Table \ref{tab:cit_reader_stats}), we conclude seminal papers are cited significantly more than literature reviews. In case of readership, we accept the null hypothesis that the distributions of reader counts of seminal and review papers are the same (that is the number of readers doesn't distinguish between the two groups). To better understand how well each metric works in distinguishing between the two groups, we use citations and readership as features in a classification experiment.

The classification experiment relies on two approaches. First, we use a leave-one-out cross-validation setup, that is we repeatedly train on all but one publication and then test the performance of the model on the publication we left out of the training. We do this for all publications in the set. However, because in some cases, due to the size of the dataset, leaving out even one publication can affect the performance of the model, we also find the performance of the ideal model, that is we train the model on all available data. This gives us an upper bound of performance.

We run three separate experiments. First, we train and test our models on all available data. This gives us an idea of how well do both metrics perform across disciplines and regardless of time. We call this the aggregate model (Section \ref{sec:model_agg}). Next, we split the data by discipline and create separate models for each discipline (Section \ref{sec:model_disc}). Finally, we split the data by publication years and create separate models for each year (Section \ref{sec:model_year}). It would be interesting to also split the data by both discipline and year, however, we weren't able to do this due to the size of the dataset, as the resulting groups would be too small for analysis.

\subsection{Aggregate model}
\label{sec:model_agg}

The model we use to classify papers based on their citation and reader counts works in the following way: if the total number of citations (or the number of readers) for a given paper is equal to or greater than a selected threshold we classify the paper as seminal, otherwise as a literature review. To do this, we use the threshold which achieves the best accuracy (which is calculated as the number of correctly classified examples divided by the number of all examples) on the training data. We find this threshold by calculating the accuracy for all thresholds in the interval $[0, max(citation\_count)]$ for the model using citation counts and $[0, max(reader\_count)]$ for the model using reader counts. If there is more than one such threshold, we use the average value of all best thresholds. For the ideal model we chose any of the best thresholds, as all will have the same performance.

Table \ref{tab:conf_mat} shows the confusion matrix for the leave-one-out cross-validation scenario using citation counts as a feature. This setup achieves an overall accuracy of 63.06\%, which represents about 10\% improvement over the baseline. All but two of the models trained in the cross-validation setup chose 51 citations as an optimal threshold (the two other thresholds were 52.4 and 52.5). The ideal model (trained on all available data) achieves the accuracy of 63.38\%. Table \ref{tab:conf_mat_readers} shows the confusion matrix obtained by using reader counts as a feature. This model achieves an overall accuracy of 42.68\%, which is about 10\% worse than the baseline. Most of the models (277) trained in the cross-validation setup chose 0 readers as the optimal threshold.
The remaining models (37) chose 2.5 readers as a threshold. The performance of the ideal model is 52.87\%, which is equal to the baseline.

\bgroup
\def\arraystretch{1.5}
\begin{table}[ht!]
	\centering
    \small
    \begin{tabular}{l|l|c|c|c}
        \multicolumn{2}{c}{} & \multicolumn{2}{c}{Predicted} & \\
        \cline{3-4}
        \multicolumn{2}{c|}{}           & Review & Seminal & \multicolumn{1}{c}{Total} \\
        \cline{2-4}
        \multirow{2}{*}{Actual}   & Review  & 19.43\% (61)  & 27.71\% (87)    & $148$ \\
        \cline{2-4}
								        & Seminal & 9.24\% (29)    & 43.63\% (137)   & $166$ \\
        \cline{2-4}
        \multicolumn{1}{c}{} & \multicolumn{1}{c}{Total} & \multicolumn{1}{c}{90} & \multicolumn{1}{c}{224} & \multicolumn{1}{c}{$314$}\\
    \end{tabular}
    \caption{Confusion matrix for predicting the class of the paper using Google Scholar citation counts.}
	\label{tab:conf_mat}
\end{table}
\egroup

\bgroup
\def\arraystretch{1.5}
\begin{table}[ht!]
	\centering
    \small
    \begin{tabular}{l|l|c|c|c}
        \multicolumn{2}{c}{} & \multicolumn{2}{c}{Predicted} & \\
        \cline{3-4}
        \multicolumn{2}{c|}{}           & Review  & Seminal & \multicolumn{1}{c}{Total} \\
        \cline{2-4}
        \multirow{2}{*}{Actual}   & Review  & 0.00\% (0)  & 47.13\% (148)    & 148 \\
        \cline{2-4}
								        & Seminal & 10.19\% (32)    & 42.68\% (134)   & 166 \\
        \cline{2-4}
        \multicolumn{1}{c}{} & \multicolumn{1}{c}{Total} & \multicolumn{1}{c}{32} & \multicolumn{1}{c}{282} & \multicolumn{1}{c}{314}\\
    \end{tabular}
    \caption{Confusion matrix for predicting the class of the paper using Mendeley reader counts.}
	\label{tab:conf_mat_readers}
\end{table}
\egroup

\subsection{Discipline based model}
\label{sec:model_disc}

This model uses discipline information to first split the papers into groups. For all separate groups we then perform the same statistical test and classification experiment using both citation and reader counts. In this case, we remove all papers labeled as ``Other''. Furthermore, we remove all subject areas which contain less than two of each type of papers, to be able to train and test the models on representatives of both seminal and review papers. The p-value is greater than 1\% for all remaining disciplines and for both citation and reader counts, which means in all cases we accept the null hypothesis of equal averages. All p-values are shown in Appendix \ref{sec:appendix_results}, Table \ref{tab:ttest_disciplines}.

The overall cross-validation accuracy is 45.28\% for citations and 42.13\% for reader counts, which is worse than the baseline (52.87\%) in both cases. We believe this is due to the fact the baseline isn't dependent on the size of the data, while in the leave-one-out cross-validation, removing even one paper can change the performance of the model. Furthermore, the baseline method ``knows'' which class is the majority class, while our model doesn't use this information. Both of these factors make it harder to outperform the baseline. The results for separate disciplines are reported in Tables \ref{tab:classif_citations_discipline} and \ref{tab:classification_readership_discipline}.

To calculate the overall accuracy, rather than counting average accuracy across all disciplines, we sum all confusion matrices and calculate the accuracy from the sum (Tables \ref{tab:conf_mat_disc} and \ref{tab:conf_mat_disc_readers}, this method is sometimes referred to as micro-averaging). The accuracy of the optimal model goes up in both cases, to 68.11\% in the case of citations and to 62.60\% in the case of readership. This shows that separating papers by discipline has the potential of improving the results.

\bgroup
\def\arraystretch{1.5}
\begin{table}[ht!]
	\centering
    \small
    \begin{tabular}{l|l|c|c|c}
        \multicolumn{2}{c}{} & \multicolumn{2}{c}{Predicted} & \\
        \cline{3-4}
        \multicolumn{2}{c|}{}           & Review  & Seminal & \multicolumn{1}{c}{Total} \\
        \cline{2-4}
        \multirow{2}{*}{Actual}   & Review  & 24.41\% (62)  & 23.62\% (60)    & $122$ \\
        \cline{2-4}
								        & Seminal & 31.10\% (79)    & 20.87\% (53)   & $132$ \\
        \cline{2-4}
        \multicolumn{1}{c}{} & \multicolumn{1}{c}{Total} & \multicolumn{1}{c}{$141$} & \multicolumn{1}{c}{$113$} & \multicolumn{1}{c}{$254$}\\
    \end{tabular}
    \caption{Overall classification results obtained from running the classification for each discipline separately, using citations as a feature.}
	\label{tab:conf_mat_disc}
\end{table}
\egroup

\bgroup
\def\arraystretch{1.5}
\begin{table}[ht!]
	\centering
    \small
    \begin{tabular}{l|l|c|c|c}
        \multicolumn{2}{c}{} & \multicolumn{2}{c}{Predicted} & \\
        \cline{3-4}
        \multicolumn{2}{c|}{}           & Review  & Seminal & \multicolumn{1}{c}{Total} \\
        \cline{2-4}
        \multirow{2}{*}{Actual}   & Review  & 17.32\% (44)  & 30.71\% (78)    & $122$ \\
        \cline{2-4}
								        & Seminal & 27.17\% (69)    & 24.80\% (63)   & $132$ \\
        \cline{2-4}
        \multicolumn{1}{c}{} & \multicolumn{1}{c}{Total} & \multicolumn{1}{c}{$113$} & \multicolumn{1}{c}{$141$} & \multicolumn{1}{c}{$254$}\\
    \end{tabular}
    \caption{Overall classification results obtained from running the classification for each discipline separately, using reader counts as a feature.}
	\label{tab:conf_mat_disc_readers}
\end{table}
\egroup

\subsection{Year based model}
\label{sec:model_year}

We perform a similar experiment as in case of disciplines also for publication years. We split the publications in the dataset into groups by the the year in which they were published and again leave out those groups which don't contain at least two papers of each type. The p-value is greater than 1\% for all publication years (\ref{tab:ttest_years}). The overall cross-validation accuracy is 55.23\% (Table \ref{tab:conf_mat_year}) for citation counts and 51.05\% (Table \ref{tab:conf_mat_year_readers}) for reader counts, which in the case of citation counts is an improvement both over the baseline (52.87\%) and over the previous model trained per discipline. The accuracy of the optimal model is 68.62\% in the case of citations and 65.27\% in the case of reader counts. The full results are reported in Tables \ref{tab:classif_citations_year} and \ref{tab:classif_readers_year}.

\bgroup
\def\arraystretch{1.5}
\begin{table}[ht!]
	\centering
    \small
    \begin{tabular}{l|l|c|c|c}
        \multicolumn{2}{c}{} & \multicolumn{2}{c}{Predicted} & \\
        \cline{3-4}
        \multicolumn{2}{c|}{}           & Review  & Seminal & \multicolumn{1}{c}{Total} \\
        \cline{2-4}
        \multirow{2}{*}{Actual}   & Review  & 39.75\% (95)  & 17.15\% (41)    & $136$ \\
        \cline{2-4}
								        & Seminal & 27.62\% (66)    & 15.48\% (37)   & $103$ \\
        \cline{2-4}
        \multicolumn{1}{c}{} & \multicolumn{1}{c}{Total} & \multicolumn{1}{c}{$161$} & \multicolumn{1}{c}{$78$} & \multicolumn{1}{c}{$239$}\\
    \end{tabular}
    \caption{Overall classification results obtained from running the classification for each year separately, using citations as a feature.}
	\label{tab:conf_mat_year}
\end{table}
\egroup

\bgroup
\def\arraystretch{1.5}
\begin{table}[ht!]
	\centering
    \small
    \begin{tabular}{l|l|c|c|c}
        \multicolumn{2}{c}{} & \multicolumn{2}{c}{Predicted} & \\
        \cline{3-4}
        \multicolumn{2}{c|}{}           & Review  & Seminal & \multicolumn{1}{c}{Total} \\
        \cline{2-4}
        \multirow{2}{*}{Actual}   & Review  & 37.66\% (90)  & 19.25\% (46)    & $136$ \\
        \cline{2-4}
								        & Seminal & 29.71\% (71)    & 13.39\% (32)   & $103$ \\
        \cline{2-4}
        \multicolumn{1}{c}{} & \multicolumn{1}{c}{Total} & \multicolumn{1}{c}{$161$} & \multicolumn{1}{c}{$78$} & \multicolumn{1}{c}{$239$}\\
    \end{tabular}
    \caption{Overall classification results obtained from running the classification for each year separately, using reader counts as a feature.}
	\label{tab:conf_mat_year_readers}
\end{table}
\egroup

\subsection{Results}
Table \ref{tab:summary} shows a summary of classification results of all three models. The year based model performs better than the discipline based model, however this might be due to the distribution of survey and seminal publications in our dataset -- as we have shown in Table \ref{tab:years}, seminal papers in our dataset are on average older than literature reviews, which makes the year based classification easier. In reality papers published in a given year will be distributed more evenly. The performance of the discipline based model should be more stable, as the distribution of seminal and survey papers across disciplines in our dataset is more even. We haven't performed a classification across both disciplines and years as due to their wide distribution we weren't able to find enough examples belonging to the same discipline and year. The aggregate model outperforms the two other models, however, we believe this might be due to the size of the dataset. The accuracy of the ideal models suggests splitting the publications both by discipline and by year has the potential of improving the results.

\bgroup
\def\arraystretch{1.5}
\begin{table}[ht!]
	\centering
    \small
    \begin{tabular}{|l|l||c|c|}
    	\hline
        \textbf{Model} & \textbf{Data} & \textbf{Accuracy} & \textbf{Ideal acc.} \\
        \hline
        \hline
        \multirow{2}{*}{\textbf{Baseline}} & Citations & - & 52.87\% \\
        \cline{2-4}
        & Readership & - & 52.87\% \\
        \hline
        \multirow{2}{*}{\textbf{Aggregate}} & Citations & 63.06\% & 63.38\% \\
        \cline{2-4}
        & Readership & 42.68\% & 52.87\% \\
        \hline
        \multirow{2}{*}{\textbf{Discipline based}} & Citations & 45.28\% & 68.11\% \\
        \cline{2-4}
        & Readership & 42.13\% & 62.60\% \\
        \hline
        \multirow{2}{*}{\textbf{Year based}} & Citations & 55.23\% & 68.62\% \\
        \cline{2-4}
        & Readership & 51.05\% & 65.27\% \\
        \hline
    \end{tabular}
    \caption{Summary of all results. Column \textit{Accuracy} shows the accuracy obtained in the leave-one-out cross-validation scenario, while column \textit{Ideal acc.} shows a theoretical upper bound of performance (an accuracy of a model trained on all available data).}
	\label{tab:summary}
\end{table}
\egroup

\section{Discussion}
\label{sec:discussion}

We believe this study is novel in two ways. Firstly, our experiments show that citation counts help in distinguishing important seminal research from literature reviews with a degree of accuracy (63\%, i.e. 10\% over the random baseline), while Mendeley reader counts don’t work better than a random baseline on this task and our dataset. There has been much discussion whether citation counts are appropriate for use in evaluation of research outputs \citep{wilsdon2015metrictide}. We have used a new approach to study this question. In addition, our contributions include the creation of a novel dataset of 314 seminal publications and literature reviews, which is publicly available. We believe this dataset will be useful in developing and evaluating new metrics. 

While our results are statistically significant ($p < 0.01$), a larger dataset would be helpful, especially for studying differences across disciplines. We believe an ``ideal'' dataset for evaluating research metrics should meet the following requirements:

\begin{itemize}
	\item \textbf{Cross-disciplinary:} A dataset containing publications from different scientific areas is important for two reasons. Firstly, publication patterns are different for each discipline, both in terms of productivity and types of outcomes (conference papers, journal papers, books, etc.). This is also important to enable detecting research which finds use outside of its domain.
    \item \textbf{Time span:} The dataset should also contain publications spanning a wider time frame. One of the reasons for this is that publication patterns are different not only across disciplines, but they keep changing also in time. Furthermore, some research publications only find use after a certain period of time, but nevertheless represent important research.
    \item \textbf{Publication types:} Different types of research publications (e.g. pure research, applied research, literature review, dataset description, etc.) provide different types of impact. This should be taken into account when developing new research metrics. For example, a publication presenting a system might not receive many citations, because it presents a final product rather than research others can build on. However, such publication might still be widely used and have a large societal or economic impact.
    \item \textbf{Peer review judgements:} Finally, to provide a reference rank for comparing the research metrics to, the dataset should contain fair and unbiased judgements provided by domain experts. These judgements should rate the publications based on an agreed set of rules and standards.
\end{itemize}

Creating such a dataset would require significant time and resources, both in terms of collecting a representative sample of publications and in terms of providing peer review judgements for these publications. While there was a recent effort to create such a dataset (Section \ref{sec:relwork}), in this case the evaluation set contained only publications from one discipline (computer science) and the peer review judgements were not shared. Providing the peer review judgements could be a common effort and an existing open peer review system could be used for this task. This would require selecting the reference publications, creating a set of rules according to which the papers in the set should be judged and ensuring fairness of the peer review. We believe our study represents the first step in the direction of an ideal evaluation set, as utilising different publication types for metrics evaluation is currently possible. While the creation of such dataset is still time-consuming, it is a less constrained task.

One limitation of our study is that we rely on the respondents' understanding of seminal publications and literature reviews. We have verified the correctness of the responses belonging to the Computer Science and Informatics subset (43 publications), as that is an area most familiar to us. To do this, we have reviewed the publication titles and abstracts. The labelling of this subset matches our understanding of seminal and review publications except in three cases, a paper ``From data mining to knowledge discovery in databases'' which was labelled as seminal and papers ``Process algebra for synchronous communication'' and ``Unifying heterogeneous and distributed information about marine species through the top level ontology MarineTLO'' which were both labelled as a literature review. For these three papers we would flip the labels. We haven't however read the full papers and so our disagreement with the respondents could be caused by not knowing the content of the papers and/or not being experts in those areas. As future work we are planning to cross-reference the data to ensure the validity of the entire dataset.



\section{Related work}
\label{sec:relwork}
The suitability of current metrics for assessing the value of research outputs has been studied extensively in literature, especially the suitability of citations, however, other indicators \citep{bornmann2014altmetrics,thelwall2015web1}, including Mendeley readership \citep{bornmann2015usefulness,thelwall2015web2}, have been studied as well. The existing studies have typically approached the question either by studying the unit of measurement itself, for instance in the case of citations by studying the motivations of scientists for choosing to reference or to not reference specific papers \citep{harwood2009interview,macroberts2010problems} (a review of studies on citing behaviour is available in \citep{bornmann2008citation}), or in the case of Mendeley readership the reasons for bookmarking specific papers \citep{mohammadi2016can}; or by studying what does a given metric represent, for example by analysing the characteristics of highly cited papers \citep{Aksnes2003,antonakis2014makes,van2014top} or by comparing the data with another metric \citep{bornmann2015does,bornmann2015people}, typically by performing a correlation analysis. Similarly as the works studying highly cited publications, we analyse whether a high number of received citations reflects the shift a paper caused (or didn't cause) in its field. Interestingly, two of the mentioned studies have found a high proportion of the top cited papers to be literature reviews \citep{Aksnes2003} or method and software descriptions \citep{van2014top}. In contrast to previous work, we concentrate on analysing how well do important (seminal) papers perform under current evaluation methods in comparison to other types of papers (literature reviews), rather than focusing on characterising highly cited papers, or understanding what do current evaluation methods measure.

A work similar to ours was the 2016 WSDM Cup Challenge \citep{wade2016wsdm}, which has also been probably the biggest effort in this area up to date. The goal of the challenge was to provide a static rank for papers contained in the Microsoft Academic Graph (MAG) dataset \citep{sinha2015overview}. The evaluation set has been built by computer science academics using pairwise judgement on a subset of the publications in MAG. We have provided an analysis of the evaluation method in our paper \citep{herrmannova2016simple}. Unfortunately this evaluation set has not been published and so this effort does not extend beyond the challenge.

Our work is also close to several recent efforts \citep{teufel2006automatic,wan2014all,zhu2015measuring,valenzuela2015identifying,pride_2017_incidental} in which the authors argue that not all citations are equal and that identifying which citations are important is necessary for better understanding of published research. Our work provides a quantitative evidence further motivating this strand of research, as we show that while using citations works to some extent for distinguishing excellent research, there is a room for improvement. As a future work we would like to test the models presented in these studies on our dataset to see whether important citations will help in distinguishing seminal research better.

\section{Conclusions}



In this paper, we have shown citation counts work 10\% better than the baseline as indicators of excellent research, while Mendeley reader counts don't work better than the baseline. Specifically, we studied how well these two metrics distinguish publications that have changed a research field from those that have not. We have performed a set of experiments using citation and reader counts to classify papers into seminal and review categories and showed that citations distinguish between these two types of papers with low to moderate accuracy (highest accuracy achieved in all experiments was 63.06\%, while our baseline model achieved 52.87\%), while reader counts don't distinguish between them at all (highest accuracy 51.05\%). We believe this shows that while citations work to some degree, additional methods, such automated methods for classifying important citations \citep{teufel2006automatic,valenzuela2015identifying,pride_2017_incidental}, may be needed.

In addition to quantifying the success rate when using citations for identifying excellent research, we also presented a novel dataset of 314 annotated seminal publications and literature reviews along with their metadata (including DOIs, titles, authors, and abstracts), which we call TrueImpactDataset. We described how this dataset was built, provided a detailed analysis of the dataset and discussed the properties an ideal dataset for validating research evaluation metrics should have. We share this dataset with the research community\footnote{http://trueimpactdataset.semantometrics.org} and hope it will be useful to others and will perhaps inspire creating a true ground truth evaluation set.




\section{Acknowledgements}

This research did not receive any specific grant from funding agencies in the public, commercial, or not-for-profit sectors.


%
\balance
\bibliographystyle{plainnat}
\bibliography{sigproc}  
%
%
\newpage
\onecolumn
\appendix
\section{Experiment results}
\label{sec:appendix_results}

\bgroup
\def\arraystretch{1.5}
\begin{table*}[!thpb]
	\centering
    \small
    \begin{tabular}{|l|r|r|r|}
    	\hline
        \textbf{Discipline} & \textbf{$p$ (citations)} & \textbf{$p$ (readership)} & \textbf{Total} \\
        \hline
        \hline
		Geography, Environmental Studies and Archaeology & 0.3404 & 0.2081 & 8 \\
        \hline
		Biological Sciences & 0.1748 & 0.4956 & 17 \\
        \hline
		Computer Science and Informatics & 0.0895 & 0.4517 & 43 \\
        \hline
		Mathematical Sciences & 0.2549 & 0.2518 & 14 \\
        \hline
		Earth Systems and Environmental Sciences & 0.1162 & 0.1645 & 18 \\
        \hline
		Business and Management Studies & 0.1191 & 0.1577 & 19 \\
        \hline
		Physics & 0.3819 & 0.1679 & 26 \\
        \hline
		Education & 0.1162 & 0.2146 & 26 \\
        \hline
		Psychology, Psychiatry and Neuroscience & 0.2443 & 0.2293 & 9 \\
        \hline
		Politics and International Studies & 0.2007 & 0.4275 & 6 \\
        \hline
		Electrical and Electronic Engineering, Metallurgy and Materials & 0.4260 & 0.3397 & 16 \\
        \hline
		Sociology & 0.4302 & 0.3955 & 7 \\
        \hline
        Classics & 0.1265 & 0.2113 & 4 \\
        \hline
		Art and Design: History, Practice and Theory & 0.2702 & 0.4565 & 5 \\
        \hline
		Social Work and Social Policy & 0.0910 & 0.3365 & 6 \\
        \hline
		Economics and Econometrics & 0.1525 & 0.3977 & 8 \\
        \hline
        General Engineering & 0.2079 & 0.1453 & 4 \\
        \hline
        Anthropology and Development Studies & 0.2920 & 0.2850 & 4 \\
        \hline
        Aeronautical, Mechanical, Chemical and Manufacturing Engineering &  0.2439 & 0.2015 &  4 \\
        \hline
        Modern Languages and Linguistics & 0.1557 & 0.1154 & 4 \\
        \hline
		Public Health, Health Services and Primary Care & 0.2056 & 0.1906 & 6 \\
        \hline
        \hline
        \textbf{Total} & - & - & 254 \\
        \hline
    \end{tabular}
    \caption{Results of independent one-tailed t-test performed using citation and readership counts on all disciplines separately.}
	\label{tab:ttest_disciplines}
\end{table*}
\egroup

\bgroup
\def\arraystretch{1.5}
\begin{sidewaystable}
	\centering
    \small
    \begin{tabular}{|l|r|r|r|r|r|r|r|r|r|}
    	\hline
        \textbf{Discipline} & \textbf{Acc.} & \textbf{Opt.} & \textbf{Base.} & \textbf{Opt. $t$} & \textbf{TN} & \textbf{TP} & \textbf{FN} & \textbf{FP} & \textbf{Total} \\
        \hline
        \hline
		Geography, Environmental Studies and Archaeology & 0.3750 & 0.7500 & 0.5000 & 41 & 2 & 1 & 3 & 2 & 8 \\
		\hline
		Biological Sciences & 0.2941 & 0.6471 & 0.5294 & 50 & 4 & 1 & 7 & 5 & 17 \\
		\hline
		Computer Science and Informatics & 0.3023 & 0.6279 & 0.5349 & 50 & 7 & 6 & 17 & 13 & 43 \\
		\hline
		Mathematical Sciences & 0.5714 & 0.6429 & 0.5714 & 14 & 1 & 7 & 1 & 5 & 14 \\
		\hline
		Earth Systems and Environmental Sciences & 0.3333 & 0.6667 & 0.5000 & 59 & 3 & 3 & 6 & 6 & 18 \\
		\hline
		Business and Management Studies & 0.4737 & 0.6842 & 0.5263 & 197 & 6 & 3 & 6 & 4 & 19 \\
		\hline
		Physics & 0.6154 & 0.6154 & 0.5000 & 916 & 12 & 4 & 9 & 1 & 26 \\
		\hline
		Education & 0.3846 & 0.6923 & 0.5769 & 19 & 3 & 7 & 8 & 8 & 26 \\
		\hline
		Psychology, Psychiatry and Neuroscience & 0.4444 & 0.6667 & 0.5556 & 31 & 1 & 3 & 2 & 3 & 9 \\
		\hline
		Politics and International Studies & 0.6667 & 0.6667 & 0.5000 & 389 & 3 & 1 & 2 & 0 & 6 \\
		\hline
		Electrical and Electronic Engineering, Metallurgy and Materials & 0.6250 & 0.6875 & 0.5000 & 50 & 5 & 5 & 3 & 3 & 16 \\
		\hline
		Sociology & 0.7143 & 0.8571 & 0.5714 & 2 & 2 & 3 & 1 & 1 & 7 \\
		\hline
		Classics & 0.7500 & 1.0000 & 0.5000 & 25 & 2 & 1 & 1 & 0 & 4 \\
		\hline
		Art and Design: History, Practice and Theory & 0.2000 & 0.6000 & 0.6000 & 0 & 0 & 1 & 2 & 2 & 5 \\
		\hline
		Social Work and Social Policy & 0.5000 & 0.8333 & 0.5000 & 17 & 2 & 1 & 2 & 1 & 6 \\
		\hline
		Economics and Econometrics & 0.6250 & 0.7500 & 0.5000 & 119 & 3 & 2 & 2 & 1 & 8 \\
		\hline
		General Engineering & 0.5000 & 0.7500 & 0.5000 & 69 & 1 & 1 & 1 & 1 & 4 \\
		\hline
		Anthropology and Development Studies & 0.0000 & 0.5000 & 0.5000 & 0 & 0 & 0 & 2 & 2 & 4 \\
		\hline
		Aeronautical, Mechanical, Chemical and Manufacturing Engineering & 0.7500 & 0.7500 & 0.5000 & 2138 & 2 & 1 & 1 & 0 & 4 \\
		\hline
		Modern Languages and Linguistics & 0.7500 & 1.0000 & 0.5000 & 38 & 2 & 1 & 1 & 0 & 4 \\
		\hline
		Public Health, Health Services and Primary Care & 0.3333 & 0.6667 & 0.5000 & 2 & 1 & 1 & 2 & 2 & 6 \\
		\hline
		\hline
		\textbf{All} & 0.4528 & 0.6811 & -  & - & 62 & 53 & 79 & 60 & 254 \\
		\hline
    \end{tabular}
    \caption{Classification results using citation counts as a feature, performed on all disciplines separately. The columns TN, TP, FN and FP show the number of true negatives (papers correctly predicted as review), true positives (papers correctly predicted as seminal), false negatives (seminal papers incorrectly predicted as review) and false positives (review papers incorrectly predicted as seminal), respectively. The column ``Opt.'' shows accuracy achieved with the optimal model and column ``Base.'' shows accuracy of the baseline model.}
	\label{tab:classif_citations_discipline}
\end{sidewaystable}
\egroup

\bgroup
\def\arraystretch{1.5}
\begin{sidewaystable}
	\centering
    \small
    \begin{tabular}{|l|r|r|r|r|r|r|r|r|r|}
    	\hline
        \textbf{Discipline} & \textbf{Acc.} & \textbf{Opt.} & \textbf{Base.} & \textbf{Opt. $t$} & \textbf{TN} & \textbf{TP} & \textbf{FN} & \textbf{FP} & \textbf{Total} \\
        \hline
        \hline
		Geography, Environmental Studies and Archaeology & 0.0000 & 0.5000 & 0.5000 & 0 & 0 & 0 & 4 & 4 & 8 \\
		\hline
		Biological Sciences & 0.4118 & 0.5882 & 0.5294 & 123 & 6 & 1 & 7 & 3 & 17 \\
		\hline
		Computer Science and Informatics & 0.3953 & 0.5349 & 0.5349 & 0 & 0 & 17 & 6 & 20 & 43 \\
		\hline
		Mathematical Sciences & 0.0714 & 0.5714 & 0.5714 & 0 & 0 & 1 & 7 & 6 & 14 \\
		\hline
		Earth Systems and Environmental Sciences & 0.7778 & 0.7778 & 0.5000 & 96 & 5 & 9 & 0 & 4 & 18 \\
		\hline
		Business and Management Studies & 0.6316 & 0.6316 & 0.5263 & 256 & 7 & 5 & 4 & 3 & 19 \\
		\hline
		Physics & 0.2308 & 0.6154 & 0.5000 & 4 & 4 & 2 & 11 & 9 & 26 \\
		\hline
		Education & 0.6154 & 0.6154 & 0.5769 & 1 & 4 & 12 & 3 & 7 & 26 \\
		\hline
		Psychology, Psychiatry and Neuroscience & 0.3333 & 0.6667 & 0.5556 & 21 & 1 & 2 & 3 & 3 & 9 \\
		\hline
		Politics and International Studies & 0.3333 & 0.6667 & 0.5000 & 1 & 1 & 1 & 2 & 2 & 6 \\
		\hline
		Electrical and Electronic Engineering, Metallurgy and Materials & 0.5000 & 0.6250 & 0.5000 & 43 & 7 & 1 & 7 & 1 & 16 \\
		\hline
		Sociology & 0.4286 & 0.7143 & 0.5714 & 40 & 1 & 2 & 2 & 2 & 7 \\
		\hline
		Classics & 0.7500 & 0.7500 & 0.5000 & 1 & 2 & 1 & 1 & 0 & 4 \\
		\hline
		Art and Design: History, Practice and Theory & 0.2000 & 0.6000 & 0.6000 & 0 & 0 & 1 & 2 & 2 & 5 \\
		\hline
		Social Work and Social Policy & 0.1667 & 0.5000 & 0.5000 & 0 & 0 & 1 & 2 & 3 & 6 \\
		\hline
		Economics and Econometrics & 0.5000 & 0.6250 & 0.5000 & 77 & 3 & 1 & 3 & 1 & 8 \\
		\hline
		General Engineering & 0.5000 & 1.0000 & 0.5000 & 82 & 1 & 1 & 1 & 1 & 4 \\
		\hline
		Anthropology and Development Studies & 0.7500 & 0.7500 & 0.5000 & 15 & 1 & 2 & 0 & 1 & 4 \\
		\hline
		Aeronautical, Mechanical, Chemical and Manufacturing Engineering & 0.0000 & 0.5000 & 0.5000 & 0 & 0 & 0 & 2 & 2 & 4 \\
		\hline
		Modern Languages and Linguistics & 0.5000 & 1.0000 & 0.5000 & 36 & 1 & 1 & 1 & 1 & 4 \\
		\hline
		Public Health, Health Services and Primary Care & 0.3333 & 0.6667 & 0.5000 & 8 & 0 & 2 & 1 & 3 & 6 \\
		\hline
		\hline
		\textbf{All} & 0.4213 & 0.6260 & -  & - & 44 & 63 & 69 & 78 & 254 \\
		\hline
    \end{tabular}
    \caption{Classification results using reader counts as a feature, performed on all disciplines separately. The columns TN, TP, FN and FP show the number of true negatives (papers correctly predicted as review), true positives (papers correctly predicted as seminal), false negatives (seminal papers incorrectly predicted as review) and false positives (review papers incorrectly predicted as seminal), respectively. The column ``Opt.'' shows accuracy achieved with the optimal model and column ``Base.'' shows accuracy of the baseline model.}
	\label{tab:classification_readership_discipline}
\end{sidewaystable}
\egroup

\bgroup
\def\arraystretch{1.5}
\begin{table*}[!thpb]
	\centering
    \small
    \begin{tabular}{|l|r|r|r|}
    	\hline
        \textbf{Year} & \textbf{$p$ (citations)} & \textbf{$p$ (readership)} & \textbf{Total} \\
        \hline
        \hline
		1999 & 0.3738 & 0.1951 & 8 \\
        \hline
		2000 & 0.1706 & 0.0555 & 10 \\
        \hline
		2001 & 0.1988 & 0.3102 & 15 \\
        \hline
		2003 & 0.1096 & 0.3459 & 9 \\
        \hline
		2004 & 0.4157 & 0.1629 & 10 \\
        \hline
		2005 & 0.2115 & 0.3178 & 17 \\
        \hline
		2006 & 0.3230 & 0.2259 & 14 \\
        \hline
		2007 & 0.1570 & 0.1482 & 15 \\
        \hline
		2008 & 0.2112 & 0.4029 & 14 \\
        \hline
		2009 & 0.1199 & 0.0531 & 11 \\
        \hline
		2010 & 0.1098 & 0.3501 & 21 \\
        \hline
		2011 & 0.2064 & 0.2207 & 18 \\
        \hline
		2012 & 0.1154 & 0.4622 & 17 \\
        \hline
		2013 & 0.4370 & 0.1918 & 19 \\
        \hline
		2014 & 0.2785 & 0.0731 & 13 \\
        \hline
		2015 & 0.4661 & 0.1684 & 11 \\
        \hline
		2016 & 0.0842 & 0.3098 & 17 \\
        \hline
        \hline
        \textbf{Total} & - & - & 239 \\
        \hline
    \end{tabular}
    \caption{Results of independent one-tailed t-test performed using citation and readership counts on all publication years separately.}
	\label{tab:ttest_years}
\end{table*}
\egroup

\bgroup
\def\arraystretch{1.5}
\begin{table}
	\centering
    \small
    \begin{tabular}{|l|r|r|r|r|r|r|r|r|r|}
    	\hline
        \textbf{Year} & \textbf{Acc.} & \textbf{Opt.} & \textbf{Base.} & \textbf{Opt. $t$} & \textbf{TN} & \textbf{TP} & \textbf{FN} & \textbf{FP} & \textbf{Total} \\
        \hline
        \hline
		1999 & 0.7500 & 0.7500 & 0.7500 & 0 & 0 & 6 & 0 & 2 & 8 \\
		\hline
		2000 & 0.6000 & 0.7000 & 0.7000 & 0 & 0 & 6 & 1 & 3 & 10 \\
		\hline
		2001 & 0.1333 & 0.6000 & 0.5333 & 3 & 1 & 1 & 7 & 6 & 15 \\
		\hline
		2003 & 0.6667 & 0.8889 & 0.5556 & 374 & 3 & 3 & 2 & 1 & 9 \\
		\hline
		2004 & 0.3000 & 0.7000 & 0.5000 & 35 & 2 & 1 & 4 & 3 & 10 \\
		\hline
		2005 & 0.4706 & 0.5882 & 0.5882 & 472 & 8 & 0 & 7 & 2 & 17 \\
		\hline
		2006 & 0.5714 & 0.5714 & 0.5714 & 1559 & 7 & 1 & 5 & 1 & 14 \\
		\hline
		2007 & 0.6667 & 0.6667 & 0.6000 & 37 & 5 & 5 & 1 & 4 & 15 \\
		\hline
		2008 & 0.4286 & 0.7143 & 0.5000 & 197 & 2 & 4 & 3 & 5 & 14 \\
		\hline
		2009 & 0.4545 & 0.5455 & 0.6364 & 214 & 5 & 0 & 4 & 2 & 11 \\
		\hline
		2010 & 0.6190 & 0.7143 & 0.5714 & 1105 & 11 & 2 & 7 & 1 & 21 \\
		\hline
		2011 & 0.5000 & 0.6667 & 0.5556 & 59 & 3 & 6 & 4 & 5 & 18 \\
		\hline
		2012 & 0.7059 & 0.7059 & 0.6471 & 633 & 11 & 1 & 5 & 0 & 17 \\
		\hline
		2013 & 0.6316 & 0.7895 & 0.7895 & 240 & 12 & 0 & 4 & 3 & 19 \\
		\hline
		2014 & 0.6923 & 0.6923 & 0.7692 & 64 & 9 & 0 & 3 & 1 & 13 \\
		\hline
		2015 & 0.6364 & 0.7273 & 0.7273 & 96 & 7 & 0 & 3 & 1 & 11 \\
		\hline
		2016 & 0.5882 & 0.7059 & 0.5882 & 2 & 9 & 1 & 6 & 1 & 17 \\
		\hline
		\hline
		\textbf{All} & 0.5523 & 0.6862 & -  & - & 95 & 37 & 66 & 41 & 239 \\
		\hline
    \end{tabular}
    \caption{Classification results using citation counts as a feature, performed on all years separately. The columns TN, TP, FN and FP show the number of true negatives (papers correctly predicted as reivew), true positives (papers correctly predicted as seminal), false negatives (seminal papers incorrectly predicted as review) and false positives (review papers incorrectly predicted as seminal), respectively. The column ``Opt.'' shows accuracy achieved with the optimal model and column ``Base.'' shows accuracy of the baseline model.}
	\label{tab:classif_citations_year}
\end{table}
\egroup

\bgroup
\def\arraystretch{1.5}
\begin{table}
	\centering
    \small
    \begin{tabular}{|l|r|r|r|r|r|r|r|r|r|}
    	\hline
        \textbf{Year} & \textbf{Acc.} & \textbf{Opt.} & \textbf{Base.} & \textbf{Opt. $t$} & \textbf{TN} & \textbf{TP} & \textbf{FN} & \textbf{FP} & \textbf{Total} \\
        \hline
        \hline
		1999 & 0.5000 & 0.7500 & 0.7500 & 0 & 0 & 4 & 2 & 2 & 8 \\
		\hline
		2000 & 0.6000 & 0.7000 & 0.7000 & 0 & 0 & 6 & 1 & 3 & 10 \\
		\hline
		2001 & 0.5333 & 0.6667 & 0.5333 & 57 & 3 & 5 & 3 & 4 & 15 \\
		\hline
		2003 & 0.2222 & 0.5556 & 0.5556 & 0 & 0 & 2 & 3 & 4 & 9 \\
		\hline
		2004 & 0.6000 & 0.6000 & 0.5000 & 15 & 3 & 3 & 2 & 2 & 10 \\
		\hline
		2005 & 0.6471 & 0.6471 & 0.5882 & 327 & 9 & 2 & 5 & 1 & 17 \\
		\hline
		2006 & 0.2143 & 0.5714 & 0.5714 & 39 & 3 & 0 & 6 & 5 & 14 \\
		\hline
		2007 & 0.2000 & 0.6000 & 0.6000 & 10 & 3 & 0 & 6 & 6 & 15 \\
		\hline
		2008 & 0.5000 & 0.5714 & 0.5000 & 2775 & 6 & 1 & 6 & 1 & 14 \\
		\hline
		2009 & 0.4545 & 0.5455 & 0.6364 & 382 & 5 & 0 & 4 & 2 & 11 \\
		\hline
		2010 & 0.5714 & 0.6190 & 0.5714 & 326 & 11 & 1 & 8 & 1 & 21 \\
		\hline
		2011 & 0.3889 & 0.6111 & 0.5556 & 1 & 2 & 5 & 5 & 6 & 18 \\
		\hline
		2012 & 0.4118 & 0.6471 & 0.6471 & 41 & 7 & 0 & 6 & 4 & 17 \\
		\hline
		2013 & 0.7895 & 0.8421 & 0.7895 & 823 & 14 & 1 & 3 & 1 & 19 \\
		\hline
		2014 & 0.6154 & 0.6923 & 0.7692 & 123 & 8 & 0 & 3 & 2 & 13 \\
		\hline
		2015 & 0.7273 & 0.8182 & 0.7273 & 1028 & 7 & 1 & 2 & 1 & 11 \\
		\hline
		2016 & 0.5882 & 0.6471 & 0.5882 & 35 & 9 & 1 & 6 & 1 & 17 \\
		\hline
		\hline
		\textbf{All} & 0.5105 & 0.6527 & -  & - & 90 & 32 & 71 & 46 & 239 \\
		\hline
    \end{tabular}
    \caption{Classification results using reader counts as a feature, performed on all years separately. The columns TN, TP, FN and FP show the number of true negatives (papers correctly predicted as review), true positives (papers correctly predicted as seminal), false negatives (seminal papers incorrectly predicted as review) and false positives (review papers incorrectly predicted as seminal), respectively. The column ``Opt.'' shows accuracy achieved with the optimal model and column ``Base.'' shows accuracy of the baseline model.}
	\label{tab:classif_readers_year}
\end{table}
\egroup

\end{document}